\UseRawInputEncoding
\documentclass[preprintnumbers,superscriptaddress,amsmath,amssymb,prd,preprint,nofootinbib]{revtex4-1}
\usepackage{amssymb}
\usepackage{mathrsfs}
\usepackage{amsthm}
\usepackage{graphicx}
\usepackage{amsfonts}
\usepackage[colorlinks,linkcolor=red,anchorcolor=blue,citecolor=green]{hyperref}
\usepackage{subfigure}
\usepackage{float}
\usepackage{enumerate}

\begin{document}	
\title{Analytical calculation of Kerr and Kerr-Ads black holes in $f(R)$ theory}
\author{Ping Li}
\email[]{Lip57120@huas.edu.cn}
\affiliation{College of Mathematics and Physics, Hunan University of Arts and Sciences, 3150 Dongting Dadao, Changde City, Hunan Province 415000, China}
\affiliation{Hunan Province Key Laboratory Integration and Optical Manufacturing Technology, 3150 Dongting Dadao, Changde City, Hunan Province 415000, China}
\author{Yong-qiang Liu}
\email[]{1000511286@smail.shnu.edu.cn}
\affiliation{Division of Mathematica and Theoretical Physics, Shanghai Normal University, 100 Guilin Road, Shanghai 200234, China}
\author{Jiang-he Yang}
\email[]{yjianghe@163.com}
\affiliation{College of Mathematics and Physics, Hunan University of Arts and Sciences, 3150 Dongting Dadao, Changde City, Hunan Province 415000, China}
\affiliation{Center for Astrophysics, Guangzhou University, 230 West Ring Road, Guangzhou, Guangdong Province 510006, China}
\author{Siwei Xu}
\email[]{xusiwei1227@163.com}
\affiliation{College of Mathematics and Physics, Hunan University of Arts and Sciences, 3150 Dongting Dadao, Changde City, Hunan Province 415000, China}
\affiliation{Hunan Province Key Laboratory Integration and Optical Manufacturing Technology, 3150 Dongting Dadao, Changde City, Hunan Province 415000, China}
\author{Xiang-hua Zhai}
\email[]{zhaixh@shnu.edu.cn}
\affiliation{Division of Mathematica and Theoretical Physics, Shanghai Normal University, 100 Guilin Road, Shanghai 200234, China}

\begin{abstract}
In this paper, we extend Chandrasekhar's method of calculating rotating black holes into $f(R)$ theory. We consider the Ricci scalar is a constant and derive the Kerr and Kerr-Ads metric by using the analytical mathematical method. Suppose that the spacetime is a 4-dimensional Riemannian manifold with a general stationary axisymmetric metric, we calculate Cartan's equation of structure and derive the Einstein tensor. In order to reduce the solving difficulty, we fix the gauge freedom to transform the metric into a more symmetric form. We solve the field equations in the two cases of the Ricci scalar $R=0$ and $R\neq 0$. In the case of $R=0$, the Ernst's equations are derived. We give the elementary solution of Ernst's equations and show the way to obtain more solutions including Kerr metric. In the case of $R\neq 0$, we reasonably assume that the solution to the equations consists of two parts: the first is Kerr part and the second is introduced by the Ricci scalar. Giving solution to the second part and combining the two parts, we obtain the Kerr-Ads metric. The calculations are carried out in a general $f(R)$ theory, indicating the Kerr and Kerr-Ads black holes exist widely in general $f(R)$ models. Furthermore, the whole solving process can be treated as a standard calculation procedure to obtain rotating black holes, which can be applied to other modified gravities.
\end{abstract}
\maketitle

\section{Introduction}
The black hole solutions in $f(R)$ gravity have been widely discussed. Due to the extension of gravitational action, there may be more possible black hole solutions in $f(R)$ theory. But this also brings more mathematical difficulties to search the black hole solutions. In order to obtain the metric with analytical functions, some special calculation methods are often necessary. In the static spherically symmetric (SSS) case, the Birkhoff theorem is no longer valid for general $f(R)$ theories. Multam\"{a}ki and Vilja showed how to construct SSS solutions in some special $f(R)$ models \cite{Multamaki:2006}. Capozziello \emph{et. al.} constructed the Noether symmetry method to calculate the solutions \cite{Capozziello2007,Capozziello2010}. Later, the Lagrange multiplier method was introduced by Sebastiani \cite{Sebastiani:2011} and Nashed \cite{Nashed:2019}. The Lagrangian multiplier method treats the Ricci scalar $R$ as a new variable and uses the Lagrangian multiplier to constrain it. In Refs. \cite{Nashed:2019,Elizalde:2020,Nashed:2021-1}, the authors obtained a new class of analytic SSS black hole solutions with a Ricci scalar $R\propto \frac{1}{r^2}$ in $f(R)=R-2\alpha\sqrt{R}$ model. This new black hole was also obtained in Ref. \cite{Li:2022} and may be considered as a spacetime corresponding to a fixed deficit angle $\delta \theta=\pi$ \cite{Li:2022-1}.

It is more difficult to derive axially symmetric solutions than SSS solutions in $f(R)$ gravity. An interesting consideration may be the generation of axially symmetric solutions from SSS solutions through a set of complex coordinate transformations. This method was first introduced by Newman and Janis in general relativity (GR) \cite{Newman:1965}. Since then, this algorithm has been widely used as a powerful tool to obtain the Kerr-like rotating metrics from the corresponding SSS metrics. For recent studies, please refer to \cite{Shao:2021,Kubiznak:2021,Kamenshchik:2023,Fernandes:2023}. Capozziello \emph{et. al.} have shown that the Newman-Janis algorithm also works in $f(R)$ gravity \cite{Capozziello2010}. Later, the radiating Kerr-Newman black hole is obtained by using the Newman-Janis algorithm in $f(R)$ gravity \cite{Ghosh:2013}. Recently, using a modified Newman-Janis algorithm, Chaturvedi obtained the rotating axisymmetric solutions in $f(R)$ gravity theories with a constant Ricci scalar \cite{Chaturvedi:2023}. However, the Newman-Janis algorithm does not depend on the Einstein field equations but directly works on the solutions. All solutions obtained by this method should be checked whether they fulfill the field equations. In quadratic gravity models, the Newman-Janis algorithm is not suitable for generating axisymmetric metrics \cite{Cadoni:2011,Hansen:2013,Ayzenberg:2015}. Thus, this method can not be used as a universal method for deriving axially symmetric solutions.

There are few papers aiming to derive the axisymmetric solutions through strict mathematical procedures. As far as we know, there are two different roads to obtain the Kerr metric using the strict analytical calculations. One of them originally came from Carter \cite{Carter}, who used an additional physical condition to simplify the line element. Suppose a test particle traveling in the black hole spacetime, the correspondence of quantum mechanics requires that the Klein-Gordon equation is separable. Therefore, the spacetime of black hole should have a special structure and the rotating metric could be written in a separable canonical form. Using the separable canonical metric form, Carter obtained the axisymmetric solutions including Kerr and Kerr-Ads solutions in Ref. \cite{Carter}.

The other road is given by Chandrasekhar \cite{Chandrasekhar1983}. In GR, Chandrasekhar used a general metric of stationary axisymmetric spacetime to derive the Kerr and Kerr-Newman solutions. After lots of variable substitutions, the Ernst's equations are obtained. As is shown in Ref.\cite{Chandrasekhar1983}, the Kerr solution is a special case of the elementary solutions of the Ernst's equations. The biggest benefit of this process is the strict analytical calculation, which also makes the method universal.

In this paper, we derive the axisymmetric solutions using the method mainly adopted from Chandrasekhar. Unlike in GR, the field equations of $f(R)$ gravity would be highly coupled and not easy to solve. Fortunately, after adopting some strategies, we do solve the field equations analytically. In the cases of $R=0$ and $R\neq 0$, we obtain Kerr and Kerr-Ads solutions, respectively. What counts is that the whole process is highly universal and can be easily generalized to derive the axisymmetric solutions in other modified gravitational theories.

This paper is organized as follows. In Sec.II, we review the vacuum field equations with a constant Ricci scalar in a general $f(R)$ theory. In Sec.III, based on a general stationary axisymmetric metric, we calculate the Cartan's structure equation and obtain the Einstein tensor. In Sec.IV, we choose a highly symmetrical gauge to fix the metric form and derive the field equations. In Sec.V, variables are transformed to the conjugate metric and some preprocess is done for solving. In Sec.VI, the Kerr metric is derived by solving the Ernst's equations. In Sec.VII, the Kerr-Ads metric is derived. The last section is a detailed conclusion.

In this paper, the Greek letters $\alpha,\beta,\gamma,...=0,1,2,3$ represent indices of the natural coordinate base, The Latin letters $a,b,c,d=0,1,2,3$ represent indices of moving frame. And the Latin letter $i,j,k=1,2,3$ represent the spatial indices.

\section{The Action}
The action of $f(R)$ theory is
\begin{equation}\label{action}
  S=\int d^4x \sqrt{-g}(f(R)+\mathcal{L}_m),
\end{equation}
where $\mathcal{L}_m$ is the Lagrangian density of matter. We set $8\pi G=1$. The field equations can be written as
\begin{equation}\label{eom}
 f_RR_{\alpha\beta}-\frac{1}{2}fg_{\alpha\beta}+(g_{\alpha\beta}\square+\nabla_{\alpha}\nabla_{\beta})f_R=T_{\alpha\beta},
\end{equation}
where $f_R$ stands for the derivative of $f$ with respect to $R$. The Einstein tensor is defined by
\begin{equation}\label{ET}
G_{\alpha\beta}=R_{\alpha\beta}-\frac{1}{2}g_{\alpha\beta}R.
\end{equation}
Using Einstein tensor, the field equation (\ref{eom}) is expressed as
\begin{equation}
 f_RG_{\alpha\beta}+\frac{g_{\alpha\beta}}{2}(Rf_R-f)+(g_{\alpha\beta}\square+\nabla_{\alpha}\nabla_{\beta})f_R=T_{\alpha\beta}.
\end{equation}
In this paper, we consider $f_R\neq 0$ and in vacuum case $T_{\alpha\beta}=0$. In order to compare with GR, we express the vacuum field equation as
\begin{equation}
 G_{\alpha\beta}+\mathcal{T}_{\alpha\beta}=0,
\end{equation}
where
\begin{align}
  \mathcal{T}_{\alpha\beta}&=\frac{g_{\alpha\beta}}{2}(R-\frac{f}{f_R})+\mathcal{K}_{\alpha\beta},\\
  \mathcal{K}_{\alpha\beta}&=\frac{1}{f_R}(g_{\alpha\beta}\square+\nabla_{\alpha}\nabla_{\beta})f_R.
\end{align}

The Bianchi identity $\nabla^{\alpha}G_{\alpha\beta}=0$ indicates $\nabla^{\alpha}\mathcal{T}_{\alpha\beta}=0$. Through some mathematical process, one obtains
\begin{equation}\label{constrainR}
  R_{,\alpha}=-\frac{2f_R^2}{ff_{RR}}\nabla^{\beta}\mathcal{K}_{\beta\alpha},
\end{equation}
where $R_{,\alpha}$ stands for the derivative of $R$ with respect to $x^{\alpha}$. The above equation (\ref{constrainR}) is an extra equation constraining $R$ in $f(R)$ theory.

Furthermore, if the Ricci scalar $R$ is a constant, one obtains $\mathcal{K}^{\alpha}{}_{\beta}=0$ immediately. Then, the constraint equation (\ref{constrainR}) is satisfied automatically. Only the diagonal terms of $\mathcal{T}_{\alpha\beta}$ are non-zero and have the same value, i.e.
\begin{equation}
 \mathcal{T}^{0}{}_{0}=\mathcal{T}^{1}{}_{1}=\mathcal{T}^{2}{}_{2}=\mathcal{T}^{3}{}_{3}=\frac{1}{2}(R-\frac{f}{f_R}).
\end{equation}
Because the trace of Einstein tensor $G^{\alpha}{}_{\alpha}$ is $-R$, one obtains $\frac{f}{f_R}=\frac{R}{2}$ if tracing the vacuum field equation. Then, we have $\frac{1}{2}(R-\frac{f}{f_R})=\frac{R}{4}$. We emphasize that the constraint $\frac{f}{f_R}=\frac{R}{2}$ is always true for constant Ricci scalar case in $f(R)$ theory. And one can't use it to rebuild the $f(R)$ model.

\section{The Einstein tensor in a stationary axisymmetric metric}
The ans\"{a}tz of stationary axisymmetric metric is chosen to be
\begin{equation}\label{Ansatz}
  ds^2=-e^{2\nu(x^2,x^3)}dt^2+e^{2\psi(x^2,x^3)}(d\varphi-q(x^2,x^3)dt)^2+e^{2\mu_2(x^2,x^3)}(dx^2)^2
  +e^{2\mu_3(x^2,x^3)}(dx^3)^2,
\end{equation}
where $(x^0,x^1,x^2,x^3)=(t,\varphi,x^2,x^3)$ are the coordinate axes. And all unknown variables $\nu,\psi,\mu_2,\mu_3,q$ are the functions of $x^2$ and $x^3$. As is shown by Chandrasekhar \cite{Chandrasekhar1983}, the metric (\ref{Ansatz}) is the most general form of stationary axisymmetric metric.

We use the method of moving frame to calculate the Einstein tensor $G_{\alpha\beta}$. We consider the spacetime is a Riemannian manifold that is torsion-free and metric compatible. Then, the geometric properties of the manifold are fully described by the frame 1-forms $\omega^a$ and the Riemann Levi-Civita (RLC) connection forms $\omega_{ab}$. They satisfy Cartan's equation of structure
\begin{align}
  \mathrm{d} \omega^{a}+\omega^{a}{}_{b}\wedge \omega^{b} & =0,\label{CartanS1} \\
 \mathrm{d}\omega^{a}{}_{b}+ \omega^{a}{}_{c}\wedge \omega^{c}{}_{b}& =\Omega^{a}{}_{b},\label{CartanS2}
\end{align}
where $\mathrm{d}$ is the exterior differential operator and $\wedge$ stands for Cartan wedge. $\Omega^{a}{}_{b}$ are curvature 2-forms. In the coordinate frame $\omega_{\alpha}=\partial_{\alpha}$, they can be written in terms of Christoffel symbols $\Gamma^{\gamma}_{\alpha\beta}$ as
\begin{align}
  \omega_{\alpha}{}_{\beta} &=\Gamma^{\alpha}_{\beta \gamma}dx^{\gamma},  \\
  \Omega^{\alpha}{}_{\beta} & =\frac{1}{2}(\Gamma^{\alpha}_{\beta \sigma,\tau}+\Gamma^{\alpha}_{\gamma \tau}\Gamma^{\gamma}_{\beta \sigma})dx^{\tau}\wedge dx^{\sigma}.
\end{align}

Based on the metric (\ref{Ansatz}), the frame 1-forms are chosen to be
\begin{equation}\label{frame}
  \omega^0 =e^{\nu}dt, \quad
  \omega^1 =e^{\psi}(d\varphi-qdt),\quad
  \omega^2=e^{\mu_{2}}dx^2,\quad
  \omega^3=e^{\mu_{3}}dx^3.
\end{equation}
These forms are orthogonal in the Minkowski metric $\eta_{ab}=diag(-1,1,1,1)$. One may also reversely express Eq. (\ref{frame}) to
\begin{equation}\label{frame1}
 dt=e^{-\nu}\omega^0,\quad
 d\varphi=q e^{-\nu}\omega^0+e^{-\psi} \omega^1,\quad
 dx^2=e^{-\mu_{2}} \omega^2,\quad
 dx^3=e^{-\mu_{3}} \omega^3.
\end{equation}
On the other hand, the RLC connection forms in Riemannian manifold satisfy
\begin{equation}\label{frame2}
 \omega^{0}{}_{i}  =\omega^{i}{}_{0}, \quad\omega^{i}{}_{j}  =-\omega^{j}{}_{i}.
\end{equation}
The above equations (\ref{frame1}) and (\ref{frame2}) are used to simplify the calculation results.

Exterior differentiating the frame (\ref{frame}), one obtains
\begin{align}
 \mathrm{d}\omega^0 & =-e^{-\mu_2}\nu_{,2}\omega^0\wedge \omega^2-e^{-\mu_3}\nu_{,3}\omega^0\wedge \omega^3, \\
  \mathrm{d}\omega^1 & =-(e^{-\mu_2}\psi_{,2}\omega^1-e^{\psi-\nu-\mu_2}q_{,2}\omega^0)\wedge \omega^2-(e^{-\mu_3}\psi_{,3}\omega^1-e^{\psi-\nu-\mu_3}q_{,3}\omega^0)\wedge \omega^3,\\
  \mathrm{d}\omega^2 & =-e^{-\mu_3}\mu_{2,3}\omega^2\wedge \omega^3, \\
  \mathrm{d}\omega^3 & =-e^{-\mu_2}\mu_{3,2}\omega^3\wedge \omega^2.
\end{align}
Comparing the results to the first Cartan's equation of structure (\ref{CartanS1}), we can extract the nonzero RLC connection forms
\begin{align}
  \omega^{0}{}_{1} &=  \omega^{1}{}_{0}=\frac{1}{2}e^{\psi-\nu-\mu_2}q_{,2}\omega^2+\frac{1}{2}e^{\psi-\nu-\mu_3}q_{,3}\omega^3,\\
  \omega^{0}{}_{2} &=\omega^{2}{}_{0}=e^{-\mu_2}\nu_{,2}\omega^0+\frac{1}{2}e^{\psi-\nu-\mu_2}q_{,2}\omega^1,\\
  \omega^{0}{}_{3} &=\omega^{3}{}_{0}=e^{-\mu_3}\nu_{,3}\omega^0+\frac{1}{2}e^{\psi-\nu-\mu_3}q_{,3}\omega^1,\\
  \omega^{2}{}_{1} &=-\omega^{1}{}_{2}=\frac{1}{2}e^{\psi-\nu-\mu_2}q_{,2}\omega^0-e^{-\mu_2}\psi_{,2}\omega^1,\\
  \omega^{3}{}_{1} &=-\omega^{1}{}_{3}=\frac{1}{2}e^{\psi-\nu-\mu_3}q_{,3}\omega^0-e^{-\mu_3}\psi_{,3}\omega^1,\\
  \omega^{2}{}_{3} &=-\omega^{3}{}_{2}=e^{-\mu_3}\mu_{2,3}\omega^2-e^{-\mu_2}\mu_{3,2}\omega^3.
\end{align}

Inserting the frame forms and the RLC connection forms into the second Cartan's equation of structure (\ref{CartanS2}), one can obtain the curvature 2-forms $\Omega^{a}{}_{b}$. The direct calculation shows the non-zero components of curvature forms are
\begin{align}
  \Omega^{0}{}_{1} & =-(e^{-2\mu_2}\psi_{,2}\nu_{,2}+e^{-2\mu_3}\psi_{,3}\nu_{,3}+\frac{1}{4}e^{2(\psi-\nu-\mu_2)}q^2_{,2}+\frac{1}{4}e^{2(\psi-\nu-\mu_3)}q^2_{,3}) \omega^0\wedge \omega^{1}\nonumber\\
  &+\frac{1}{2}e^{-\mu_2-\mu_3}((e^{\psi-\nu}q_{,3})_{,2}-e^{\psi-\nu}q_{,2})_{,3})\omega^{2}\wedge\omega^{3},\\
  \Omega^{0}{}_{2} & =(e^{-\nu-\mu_2}(e^{\nu-\mu_2}\nu_{,2})_{,2}-\frac{3}{4}e^{2(\psi-\nu-\mu_2)}q_{,2}^2+e^{-2\mu_3}\nu_{,3}\mu_{2,3})\omega^{2}\wedge\omega^{0}\nonumber\\
  &+(e^{-\nu-\mu_3}(e^{\nu-\mu_2}\nu_{,2})_{,3}-\frac{3}{4}e^{2(\psi-\nu)-\mu_2-\mu_3}q_{,2}q_{,3}-e^{-\mu_2-\mu_3}\nu_{,3}\mu_{3,2})\omega^{3}\wedge\omega^{0}\nonumber\\
  &-\frac{1}{2}(e^{-\psi-\mu_2}( e^{2\psi-\nu-\mu_2}q_{,2})_{,2}+e^{\psi-\nu-2\mu_2}q_{,2}\psi_{,2}+e^{\psi-\nu-2\mu_3}q_{,3}\mu_{2,3})\omega^{1}\wedge\omega^{2}\nonumber\\
  &-\frac{1}{2}(e^{-\psi-\mu_3}( e^{2\psi-\nu-\mu_2}q_{,2})_{,3}+e^{\psi-\nu-\mu_2-\mu_3}q_{,3}\psi_{,2}-e^{\psi-\nu-\mu_2-\mu_3}q_{,3}\mu_{3,2})\omega^{1}\wedge\omega^{3},\\
  \Omega^{0}{}_{3} & =(e^{-\nu-\mu_3}(e^{\nu-\mu_3}\nu_{,3})_{,3}-\frac{3}{4}e^{2(\psi-\nu-\mu_3)}q_{,3}^2+e^{-2\mu_2}\nu_{,2}\mu_{3,2})\omega^{3}\wedge\omega^{0}\nonumber\\
  &
  +(e^{-\nu-\mu_2}(e^{\nu-\mu_3}\nu_{,3})_{,2}-\frac{3}{4}e^{2(\psi-\nu)-\mu_2-\mu_3}q_{,2}q_{,3}
  -e^{-\mu_2-\mu_3}\nu_{,2}\mu_{2,3})\omega^{2}\wedge\omega^{0}\nonumber\\
  &-\frac{1}{2}(e^{-\psi-\mu_2}( e^{2\psi-\nu-\mu_3}q_{,3})_{,2}+e^{\psi-\nu-\mu_2-\mu_3}q_{,2}\psi_{,3}-e^{\psi-\nu-\mu_2-\mu_3}q_{,2}\mu_{2,3})\omega^{1}\wedge\omega^{2}\nonumber\\
  &-\frac{1}{2}(e^{-\psi-\mu_3}( e^{2\psi-\nu-\mu_3}q_{,3})_{,3}+e^{\psi-\nu-2\mu_3}q_{,3}\psi_{,3}+e^{\psi-\nu-2\mu_2}q_{,2}\mu_{3,2})\omega^{1}\wedge\omega^{3},\\
  \Omega^{1}{}_{2} & =-\frac{1}{2}(e^{-\nu-\mu_2}(e^{\psi-\mu_2}q_{,2})_{,2}+2e^{\psi-\nu-2\mu_2}q_{,2}\psi_{,2}-e^{\psi-\nu-2\mu_2}\nu_{,2}q_{,2}+e^{\psi-\nu-2\mu_3}q_{,3}\mu_{2,3})\omega^{2}\wedge\omega^{0}\nonumber\\
  &-\frac{1}{2}(e^{-\nu-\mu_3}(e^{\psi-\mu_2}q_{,2})_{,3}+2e^{\psi-\nu-\mu_2-\mu_3}q_{,3}\psi_{,2}-e^{\psi-\nu-\mu_2-\mu_3}\nu_{,2}q_{,3}+e^{\psi-\nu-\mu_2-\mu_3}q_{,3}\mu_{3,2})\omega^{3}\wedge\omega^{0}\nonumber\\
  &+(e^{-\psi-\mu_2}(e^{\psi-\mu_2}\psi_{,2})_{,2}+\frac{1}{4} e^{2(\psi-\nu-\mu_2)}q_{,2}^2+e^{-2\mu_3}\psi_{,3}\mu_{2,3})\omega^{2}\wedge\omega^{1}\nonumber\\
  &+(e^{-\psi-\mu_3}(e^{\psi-\mu_2}\psi_{,2})_{,3}+\frac{1}{4} e^{2(\psi-\nu)-\mu_2-\mu_3}q_{,2}q_{,3}-e^{-\mu_2-\mu_3}\psi_{,3}\mu_{3,2})\omega^{3}\wedge\omega^{1},\\
  \Omega^{1}{}_{3} & =-\frac{1}{2}(e^{-\nu-\mu_2}(e^{\psi-\mu_3}q_{,3})_{,2}+2e^{\psi-\nu-\mu_2-\mu_3}q_{,2}\psi_{,3}-e^{\psi-\nu-\mu_2-\mu_3}\nu_{,3}q_{,2}+e^{\psi-\nu-\mu_2-\mu_3}q_{,2}\mu_{2,3})\omega^{2}\wedge\omega^{0}\nonumber\\
  &-\frac{1}{2}(e^{-\nu-\mu_3}(e^{\psi-\mu_3}q_{,3})_{,3}+2e^{\psi-\nu-2\mu_3}q_{,3}\psi_{,3}-e^{\psi-\nu-2\mu_2}\nu_{,3}q_{,3}+e^{\psi-\nu-2\mu_2}q_{,2}\mu_{3,2})\omega^{3}\wedge\omega^{0}\nonumber\\
  &+(e^{-\psi-\mu_2}(e^{\psi-\mu_3}\psi_{,3})_{,2}+\frac{1}{4} e^{2(\psi-\nu)-\mu_2-\mu_3}q_{,2}q_{,3}-e^{-\mu_2-\mu_3}\psi_{,2}\mu_{2,3})\omega^{2}\wedge\omega^{1}\nonumber\\
  &+(e^{-\psi-\mu_3}(e^{\psi-\mu_3}\psi_{,3})_{,3}+\frac{1}{4} e^{2(\psi-\nu-\mu_3)}q_{,3}^2-e^{-2\mu_2}\psi_{,2}\mu_{3,2})\omega^{3}\wedge\omega^{1},\\
  \Omega^{2}{}_{3} & =\frac{1}{2}e^{\psi-\nu-\mu_2-\mu_3}((\psi-\nu)_{,3}q_{,2}-(\psi-\nu)_{,2}q_{,3})\omega^{0}\wedge\omega^{1}\nonumber\\
  &-
  e^{-\mu_2-\mu_3}((e^{\mu_2-\mu_3}\mu_{2,3})_{,3}+(e^{\mu_3-\mu_2}\mu_{3,2})_{,2})\omega^{2}\wedge\omega^{3}.
\end{align}

The Ricci tensor $R_{ab}$ is defined by the inner product of the curvature 2-form $\Omega_{ab}$ and the dual vectors $e_{a}$
\begin{equation}\label{RicciT}
  R_{ab}=\Omega^{c}{}_{a}(e_c,e_b)\equiv\langle\Omega^{c}{}_{a};e_c,e_b\rangle,
\end{equation}
where $\langle;\rangle$ is the inner product. A normalized orthometric basis $\omega^a$ and its dual basis $e_{a}$ satisfy
\begin{equation}
  \langle\omega^a;e_b\rangle=\delta^a_b.
\end{equation}
Note that the inner product can also operate on the $n$-form and $n$ dual vectors. For the 2-form basis $\omega^a\wedge\omega^b$ and two dual basis vectors $e_c,e_d$, we have
\begin{equation}
\langle\omega^a\wedge\omega^b;e_c,e_d\rangle=\delta^{ab}_{cd},
\end{equation}
where $\delta^{ab}_{cd}$ is the generalized Kronecker-$\delta$ function.

Calculating Eq. (\ref{RicciT}), one obtains $R_{ab}$. Since the Ricci tensor $R^{\alpha}{}_{\beta}$ and $R^a{}_b$ have obvious physical significance, they have the same values whether they are in the moving frame or in the natural coordinates. So does the Ricci scalar $R$. Then, inserting the results into Eq. (\ref{ET}), we obtain $G^{\alpha}{}_{\beta}$. The calculation is straightforward and tedious. For more information about this calculation, we recommend the book \cite{Chandrasekhar1983}. At last, we obtain nonzero components of $G^{\alpha}{}_{\beta}$ as
\begin{align}
 G^{0}{}_{0}  =&e^{-2\mu_2}\left(\psi_{,2,2}+\psi^2_{,2}+\psi_{,2}(\mu_3-\mu_2)_{,2}\right)
 +e^{-2\mu_3}\left(\psi_{,3,3}+\psi^2_{,3}+\psi_{,3}(\mu_2-\mu_3)_{,3}\right)\nonumber\\
 &+e^{-\mu_2-\mu_3}\left((e^{\mu_2-\mu_3}\mu_{2,3})_{,3}+(e^{\mu_3-\mu_2}\mu_{3,2})_{,2}
\right) +\frac{1}{4}e^{2(\psi-\nu-\mu_2)}q_{,2}^2+\frac{1}{4}e^{2(\psi-\nu-\mu_3)}q_{,3}^2, \label{G00}\\
 G^{1}{}_{1} = &e^{-2\mu_2}\left(\nu_{,2,2}+\nu^2_{,2}+\nu_{,2}(\mu_3-\mu_2)_{,2}\right)
 +e^{-2\mu_3}\left(\nu_{,3,3}+\nu^2_{,3}+\nu_{,3}(\mu_2-\mu_3)_{,3}\right)\nonumber\\
 &+e^{-\mu_2-\mu_3}\left((e^{\mu_2-\mu_3}\mu_{2,3})_{,3}+(e^{\mu_3-\mu_2}\mu_{3,2})_{,2}
\right) -\frac{3}{4}e^{2(\psi-\nu-\mu_2)}q_{,2}^2-\frac{3}{4}e^{2(\psi-\nu-\mu_3)}q_{,3}^2, \label{G11}\\
 G^{2}{}_{2}=& e^{-2\mu_2}\left(\psi_{,2}\nu_{,2}+(\psi+\nu)_{,2}\mu_{3,2}\right)
 +e^{-2\mu_3}\big((\psi+\nu)_{,3,3}+\nu^2_{,3}+\psi^2_{,3}+\psi_{,3}\nu_{,3}\nonumber\\
&-(\psi+\nu)_{,3}\mu_{3,3}\big)
 +\frac{1}{4}e^{2(\psi-\nu-\mu_2)}q_{,2}^2-\frac{1}{4}e^{2(\psi-\nu-\mu_3)}q_{,3}^2, \label{G22}\\
  G^{3}{}_{3}=& e^{-2\mu_3}\left(\psi_{,3}\nu_{,3}+(\psi+\nu)_{,3}\mu_{2,3}\right)
 +e^{-2\mu_2}\big((\psi+\nu)_{,2,2}+\nu^2_{,2}+\psi^2_{,2}+\psi_{,2}\nu_{,2}\nonumber\\
&-(\psi+\nu)_{,2}\mu_{2,2}\big)
 -\frac{1}{4}e^{2(\psi-\nu-\mu_2)}q_{,2}^2+\frac{1}{4}e^{2(\psi-\nu-\mu_3)}q_{,3}^2, \label{G33}\\
 G^{0}{}_{1}=&-\frac{1}{2}e^{-(2\psi+\mu_2+\mu_3)}\left((e^{3\psi-\nu+\mu_3-\mu_2}q_{,2})_{,2}
 + (e^{3\psi-\nu+\mu_2-\mu_3}q_{,3})_{,3}\right),\label{G01}\\
 G^{2}{}_{3}=&-e^{-\mu_2-\mu_3}\left((\psi+\nu)_{,2,3}+\nu_{,2}\nu_{,3}+\psi_{,2}\psi_{,3}-(\psi+\nu)_{,2}\mu_{2,3}-(\psi+\nu)_{,3}\mu_{3,2}\right)
 \nonumber\\&+\frac{1}{2}e^{2(\psi-\nu)-\mu_2-\mu_3}q_{,2}q_{,3}.\label{G23}
\end{align}
And the Ricci scalar $R$ is expressed as
\begin{align}
 R=&-2e^{-2\mu_2}\left((\psi+\nu)_{,2,2}+\psi^2_{,2}+\nu^2_{,2}+\psi_{,2}\nu_{,2}
 +(\psi+\nu)_{,2}(\mu_3-\mu_2)_{,2} \right)\nonumber\\
 -&2e^{-2\mu_3}\left((\psi+\nu)_{,3,3}+\psi^2_{,3}+\nu^2_{,3}+\psi_{,3}\nu_{,3}
 +(\psi+\nu)_{,3}(\mu_2-\mu_3)_{,3} \right)\nonumber\\
 -&2e^{-\mu_2-\mu_3}\left((e^{\mu_2-\mu_3}\mu_{2,3})_{,3}+(e^{\mu_3-\mu_2}\mu_{3,2})_{,2} \right)+\frac{1}{2}e^{2(\psi-\nu-\mu_2)}q_{,2}^2+\frac{1}{2}e^{2(\psi-\nu-\mu_3)}q_{,3}^2.
 \label{Ricci}
\end{align}

\section{The Field equations}
Similar to Ref.\cite{Chandrasekhar1983}, we consider the following equations
\begin{align}
 G^2{}_2+G^3{}_3 & =-\frac{R}{2},\label{EinR1} \\
G^2{}_2-G^3{}_3 & =0,\label{EinR11}\\
G^1{}_1+G^0{}_0 & =-\frac{R}{2},\label{EinR21}\\
  G^1{}_1-G^0{}_0 & =0,\label{EinR2}\\
  G^0{}_1&=0.\label{EinR3}
\end{align}
Since the Bianchi identity $\nabla^{\alpha}G_{\alpha\beta}=0$ is satisfied, only 5 components of Einstein equations are independent. The direct calculation of (\ref{EinR1})-(\ref{EinR3}) gives
\begin{align}
 (e^{\psi+\nu+\mu_3-\mu_2}(\psi+\nu)_{,2})_{,2}+(e^{\psi+\nu+\mu_2-\mu_3}(\psi+\nu)_{,3})_{,3}
 &=-\frac{R}{2}e^{\psi+\nu+\mu_2+\mu_3},\label{2P3}\\
 e^{2\mu_3}(e^{\psi+\nu-\mu_2-\mu_3}(\psi+\nu)_{,2})_{,2}-e^{2\mu_2}(e^{\psi+\nu-\mu_2-\mu_3}(\psi+\nu)_{,3})_{,3}
 &-2e^{\psi+\nu+\mu_3-\mu_2}\psi_{,2}\nu_{,2}\nonumber\\
 +2e^{\psi+\nu+\mu_2-\mu_3}\psi_{,3}\nu_{,3}-\frac{1}{2}(e^{3\psi-\nu+\mu_3-\mu_2}q_{,2}^2-e^{3\psi-\nu+\mu_2-\mu_3}q_{,3}^2)&=0,\label{2M3}\\
 (e^{\psi+\nu+\mu_3-\mu_2}(\psi+\nu)_{,2})_{,2}+(e^{\psi+\nu+\mu_2-\mu_3}(\psi+\nu)_{,3})_{,3}&-2e^{\psi+\nu+\mu_3-\mu_2}\psi_{,2}\nu_{,2}\nonumber\\
 -2e^{\psi+\nu+\mu_2-\mu_3}\psi_{,3}\nu_{,3}+2e^{\psi+\nu}((e^{\mu_3-\mu_2}\mu_{3,2})_{,2}+&(e^{\mu_2-\mu_3}\mu_{2,3})_{,3})\nonumber\\
 -\frac{1}{2}(e^{3\psi-\nu+\mu_3-\mu_2}q_{,2}^2+e^{3\psi-\nu+\mu_2-\mu_3}q_{,3}^2)&=-\frac{R}{2}e^{\psi+\nu+\mu_2+\mu_3},\label{0P1}\\
 (e^{\psi+\nu+\mu_3-\mu_2}(\nu-\psi)_{,2})_{,2}+(e^{\psi+\nu+\mu_2-\mu_3}(\nu-\psi)_{,3})_{,3}
 &=e^{3\psi-\nu+\mu_3-\mu_2}q_{,2}^2+e^{3\psi-\nu+\mu_2-\mu_3}q_{,3}^2,\label{0M1}\\
 (e^{3\psi-\nu+\mu_3-\mu_2}q_{,2})_{,2}
 + (e^{3\psi-\nu+\mu_2-\mu_3}q_{,3})_{,3}&=0\label{E01}.
\end{align}
The system of equations (\ref{2P3})-(\ref{E01}) is difficult to solve. In order to obtain the solution, we need to make substitution of variables and transform the metric (\ref{Ansatz}) to a more symmetric form.

There is a gauge freedom in the plane of $(x^2,x^3)$ that can rotate the plane. In some special position, the metric can be transformed to
\begin{equation}
  e^{2\mu_2(x^2,x^3)}(dx^2)^2
  +e^{2\mu_3(x^2,x^3)}(dx^3)^2=\rho^2(r,\mu)\left(\frac{dr^2}{\Delta_r(r)}+\frac{d\mu^2}{\Delta_{\mu}(\mu)}\right),
\end{equation}
where the new coordinates $(r,\mu)$ are related to $(x^2,x^3)$ by the relationships
\begin{align}
  e^{2\mu_2(x^2,x^3)} &=\frac{\rho^2(r,\mu)}{\Delta_r} ,\\
  e^{2\mu_3(x^2,x^3)} & =\frac{\rho^2(r,\mu)}{\Delta_{\mu}}.
\end{align}
In the Boyer-Lindquist coordinate system, $r$ is the radial coordinate and $\mu=\cos\theta$. The function $\Delta_r$ depends only on $r$ and $\Delta_{\mu}$ depends only on $\mu$. In order to solve the field equations, a highly symmetrical metric form is required. Similar to $(x^2,x^3)$, there is also a gauge freedom in the plane of $(t,\varphi)$. According to \cite{Chandrasekhar1983}, this gauge freedom is chosen as
\begin{equation}\label{CForm}
 -e^{2\nu}dt^2+e^{2\psi}(d\varphi-qdt)^2=e^{\kappa}\left(-\chi dt^2+\frac{1}{\chi} (d\varphi-qdt)^2\right),
\end{equation}
where $\kappa=\kappa(r,\mu)$. More precisely, we expect $e^{\kappa}$ is a separable variable, i.e. $e^{\kappa}=h(r)l(\mu)$. As is shown in \cite{Chandrasekhar1983}, $h(r)=\sqrt{\Delta_r}$. For symmetry reasons, we choose $l(\mu)=\sqrt{\Delta_{\mu}}$. Therefore, after the gauge freedoms are fixed, the metric (\ref{Ansatz}) has a form
\begin{equation}\label{CForm}
  ds^2=\sqrt{\Delta_r\Delta_{\mu}}\left(-\chi dt^2+\frac{1}{\chi} (d\varphi-qdt)^2\right)+\rho^2\left(\frac{dr^2}{\Delta_r}+\frac{d\mu^2}{\Delta_{\mu}}\right),
\end{equation}
where $\chi=\chi(r,\mu),q=q(r,\mu)$. Now the unknown variables turn into $\chi,q,\rho^2,\Delta_r,\Delta_{\mu}$. Comparing (\ref{CForm}) with (\ref{Ansatz}), we have
\begin{equation}
   e^{2\nu}  =\chi\sqrt{\Delta_r\Delta_{\mu}},~~e^{2\psi}  =\frac{\sqrt{\Delta_r\Delta_{\mu}}}{\chi},~~e^{2\mu_2}=\frac{\rho^2}{\Delta_r},~~ e^{2\mu_3}=\frac{\rho^2}{\Delta_{\mu}}.
\end{equation}
Plugging these expressions into (\ref{2P3})-(\ref{E01}), we obtain
\begin{align}
  \Delta_{r,rr}+ \Delta_{\mu,\mu\mu} &=-R\rho^2,\label{Eq1}  \\
  \Delta_{r,rr}- \Delta_{\mu,\mu\mu}-\frac{1}{4}\frac{\Delta_{r,r}^2}{\Delta_{r}}+\frac{1}{4}\frac{\Delta_{\mu,\mu}^2}{\Delta_{\mu}}-\Delta_{r,r}(\ln\rho^2)_{,r}+&
  \Delta_{\mu,\mu}(\ln\rho^2)_{,\mu}\nonumber\\
  +\frac{\Delta_{r}}{\chi^2}(\chi^2_{,r}-q^2_{,r})-\frac{\Delta_{\mu}}{\chi^2}(\chi^2_{,\mu}-q^2_{,\mu})&=0,\label{Eq11}\\
  \Delta_{r,rr}+ \Delta_{\mu,\mu\mu}-\frac{1}{4}\frac{\Delta_{r,r}^2}{\Delta_{r}}-\frac{1}{4}\frac{\Delta_{\mu,\mu}^2}{\Delta_{\mu}}
  +2\Delta_{r}(\ln \rho^2)_{,rr}+&2\Delta_{\mu}(\ln \rho^2)_{,\mu\mu}\nonumber\\
  +\Delta_{r,r}(\ln \rho^2)_{,r}+\Delta_{\mu,\mu}(\ln \rho^2)_{,\mu}+\frac{\Delta_{r}}{\chi^2}(\chi^2_{,r}-q^2_{,r})+\frac{\Delta_{\mu}}{\chi^2}(\chi^2_{,\mu}-q^2_{,\mu})&=-R\rho^2,\label{Eq21}\\
  \frac{1}{\chi^2}(\Delta_r q_{,r}^2+\Delta_{\mu} q_{,\mu}^2)-\left(\frac{\Delta_r}{\chi}\chi_{,r}\right)_{,r}-\left(\frac{\Delta_{\mu}}{\chi}\chi_{,\mu}\right)_{,\mu} & =0,\label{Eq2}\\
  \left(\frac{\Delta_r}{\chi^2}q_{,r}\right)_{,r}+\left(\frac{\Delta_{\mu}}{\chi^2}q_{,\mu}\right)_{,\mu}&=0.\label{Eq3}
\end{align}
The above Eqs.(\ref{Eq1})-(\ref{Eq3}) determine all five unknown variables $\chi,q,\rho^2,\Delta_r,\Delta_{\mu}$. And we find $r$ and $\mu$ always appear in pairs. This is because we use the metric (\ref{CForm}) to simplify the field equations and this is also a key point to obtain the solutions.

\section{Conjugate metric}
The forms of Eqs. (\ref{Eq1}), (\ref{Eq2}) and (\ref{Eq3}) are relatively simple and two variables $r$ and $\mu$ always appear symmetrically, which is a breakthrough for solving the field equations. In the special case $R=0$, Eqs. (\ref{Eq1}), (\ref{Eq2}) and (\ref{Eq3}) contain four unknowns $\chi,q$ and $\Delta_r,\Delta_{\mu}$.
The paired appearance of $r$ and $\mu$ indicates that the dependence of $\Delta_r$'s configuration on $r$ is similar to that of $\Delta_{\mu}$ on $\mu$. In this sense, simultaneously solving Eqs. (\ref{Eq1}), (\ref{Eq2}) and (\ref{Eq3}) is possible. However, if $R$ is a non-zero constant, Eqs. (\ref{Eq1}), (\ref{Eq2}) and (\ref{Eq3}) contain an extra unknown function $\rho^2$. Strictly speaking, in the non-zero Ricci scalar case, the closed system of field equations contain all five Eqs. (\ref{Eq1})-(\ref{Eq3}), which brings great difficulty to do analytical solution. To avoid this situation, we presuppose the form of $\rho^2$ to keep the system of Eqs. (\ref{Eq1}), (\ref{Eq2}) and (\ref{Eq3}) remain closed. In this section, we do some pre-processing on Eqs. (\ref{Eq2}) and (\ref{Eq3}).

Similar to Ref. \cite{Chandrasekhar1983}, we use new variables to re-express Eqs. (\ref{Eq2}) and (\ref{Eq3}). Defining
\begin{equation}
  \Phi_{,r}=\frac{\Delta_{\mu}}{\chi^2}q_{,\mu},~~\Phi_{,\mu}=-\frac{\Delta_r}{\chi^2}q_{,r},
\end{equation}
one can rewrite Eq. (\ref{Eq3}) as $\Phi_{,\mu,r}=\Phi_{,r,\mu}$. At the same time, since $q_{,r,\mu}=q_{,\mu,r}$, the variable $\Phi$ satisfies
\begin{equation}\label{Eq4}
 \left(\frac{\chi^2}{\Delta_{\mu}} \Phi_{,r}\right)_{,r}+\left(\frac{\chi^2}{\Delta_r}\Phi_{,\mu}\right)_{,\mu}=0.
\end{equation}
Furthermore, letting
\begin{equation}
  \Psi=\frac{\sqrt{\Delta_r\Delta_{\mu}}}{\chi},
\end{equation}
we can reexpress Eqs. (\ref{Eq2}) and (\ref{Eq4}) as
\begin{align}
 \Psi[(\Delta_r \Psi_{,r})_{,r}+(\Delta_{\mu} \Psi_{,\mu})_{,\mu}]&= \Delta_r[\Psi_{,r}^{2}-\Phi_{,r}^2]+ \Delta_{\mu}[\Psi_{,\mu}^{2}-\Phi_{,\mu}^2]
 +\frac{\Psi^2}{2}(\Delta_{r,rr}+\Delta_{\mu,\mu\mu}),\label{Ernst1}\\
 \Psi[(\Delta_r \Phi_{,r})_{,r}+(\Delta_{\mu} \Phi_{,\mu})_{,\mu}]&=2\Delta_r\Phi_{,r}\Psi_{,r}+2\Delta_{\mu}\Phi_{,\mu}\Psi_{,\mu}.\label{Ernst2}
\end{align}

In order to give a simple solving process, we need to solve Eqs. (\ref{Ernst1}) and (\ref{Ernst2}) in a conjugate metric. With the coordinate transformation
\begin{equation}
  t\rightarrow i\varphi,~~~\varphi\rightarrow -it,
\end{equation}
one can make the metric transformation and obtain the conjugate metric as
\begin{align}
  \chi dt^2-\frac{1}{\chi}(d\varphi-q dt)^2\rightarrow & \chi dt^2+\frac{2q}{\chi}dtd\varphi-\frac{\chi^2-q^2}{\chi}d\varphi^2 \nonumber \\
   =&\tilde{\chi} dt^2-\frac{1}{\tilde{\chi}}(d\varphi-\tilde{q} dt)^2,\nonumber
\end{align}
where
\begin{equation}
  \tilde{\chi}=\frac{\chi}{\chi^2-q^2},~~~\tilde{q}=\frac{q}{\chi^2-q^2}.\nonumber
\end{equation}
Similarly, one can also define
\begin{align}
  \tilde{\Psi} & =\frac{\sqrt{\Delta_{r}\Delta_{\mu}}}{\tilde{\chi}}, \\
   \tilde{\Phi}_{,r}&=\frac{\Delta_{\mu}}{\tilde{\chi}^2}\tilde{q}_{,\mu},\\
  \tilde{ \Phi}_{,\mu}&=-\frac{\Delta_r}{\tilde{\chi}^2}\tilde{q}_{,r}.
\end{align}
The new variables $\tilde{\Psi}$ and $\tilde{\Phi}$ also satisfy Eqs. (\ref{Ernst1}) and (\ref{Ernst2}).

\section{The derivation of Kerr metric}
We solve Eqs. (\ref{Eq1}), (\ref{Eq2}) and (\ref{Eq3}) simultaneously in the two cases of $R=0$ and $R\neq 0$. In this section, we give the standardized solving procedure in the case of $R=0$. The main process is rooted from \cite{Chandrasekhar1983}.

In the case of $R=0$, we denote
\begin{equation}
 \Delta_{r}= \Delta_{r}^{K},~\Delta_{\mu}=\Delta_{\mu}^{K},~\tilde{\Phi}=\tilde{\Phi}^{K},~\tilde{\Psi}=\tilde{\Psi}^{K}.\nonumber
\end{equation}
Eqs. (\ref{Eq2}) and (\ref{Eq3}) reduce to
\begin{align}
  \tilde{\Psi}^K[(\Delta^K_r \tilde{\Psi}^K_{,r})_{,r}+(\Delta^K_{\mu} \tilde{\Psi}^K_{,\mu})_{,\mu}]&= \Delta^K_r[(\tilde{\Psi}_{,r}^{K})^2-(\tilde{\Phi}_{,r}^{K})^2]+ \Delta^2_{\mu}[(\tilde{\Psi}_{,\mu}^{K})^2-(\tilde{\Phi}_{,\mu}^{K})^2],\label{StK2}\\
 \tilde{\Psi}^K[(\Delta^K_r \tilde{\Phi}^K_{,r})_{,r}+(\Delta^K_{\mu} \tilde{\Phi}^K_{,\mu})_{,\mu}]&=2\Delta^K_r\tilde{\Phi}^K_{,r}\tilde{\Psi}^K_{,r}+2\Delta^K_{\mu}\tilde{\Phi}^K_{,\mu}\tilde{\Psi}^K_{,\mu}.\label{StK3}
\end{align}
In the conjugate metric, Eq. (\ref{Eq1}) remains the same. The shortest and simplest route to the Kerr metric is to give the solution of Ernst's equations. In order to obtain the Ernst's equations, we further express $\tilde{\Psi}^K$ as the real part and $\tilde{\Phi}^K$ as the imaginary part of a complex function $Z$, i.e.
\begin{equation}
 Z=\tilde{\Psi}^K+i\tilde{\Phi}^K.
\end{equation}
Then, Eqs. (\ref{StK2}) and (\ref{StK3}) can be combined into
\begin{equation}\label{Ernst3}
\mathrm{ Re}(Z)[(\Delta^K_r Z_{,r})_{,r}+(\Delta^K_{\mu} Z_{,\mu})_{,\mu}]=\Delta^K_rZ_{,r}^2+\Delta^K_{\mu}Z_{,\mu}^2,
\end{equation}
where $\mathrm{Re}(Z)$ is the real part of $Z$. One can prove that if $Z$ is the solution of Eq. (\ref{Ernst3}), $\frac{Z}{1+i C Z}$ is also a solution, where $C$ is a real constant. The transformation $Z\rightarrow \frac{Z}{1+i C Z}$ is called the Ehler's transformation. By the transformation
\begin{equation}
Z=-\frac{1+E}{1-E},
\end{equation}
one can rewrite the above Eq. (\ref{Ernst3}) to
\begin{equation}\label{Ernst}
(1-EE^{*})[(\Delta^K_r E_{,r})_{,r}+(\Delta^K_{\mu} E_{,\mu})_{,\mu}]=-2E^{*}[\Delta^K_rE_{,r}^2+\Delta^K_{\mu}E_{,\mu}^2].
\end{equation}
Eqs. (\ref{Eq1}) and (\ref{Ernst}) are called the Ernst's equations.

As is said before, the dependence of $\Delta^K_r$'s configuration on $r$ is similar to that of $\Delta^K_{\mu}$ on $\mu$. It is not difficult to show that $\Delta^K_{r}(r)$ is a quadratic function of $r$. Similarly, $\Delta^K_{\mu}(\mu)$ is a quadratic function of $\mu$. Thus, the Ernst's equations permit the elementary solution
\begin{align}
   \Delta^K_{r}&=r^2-1,  \label{SolK1}\\
   \Delta^K_{\mu}&=1-\mu^2,\label{SolK2}\\
   E&=-C_r r-i C_{\mu} \mu,
\end{align}
where $C_r^2+C_{\mu}^2=1$ and $C_r,C_{\mu}$ are real constants. Then, the corresponding solutions for $\tilde{\Psi}^K,\tilde{\Phi}^K$ are
\begin{align}
   \tilde{\Psi}^K&=\frac{C_r^2\Delta^K_{r}-C_{\mu}^2\Delta^K_{\mu}}{(C_rr+1)^2+C_{\mu}^2\mu^2}, \label{SolK3} \\
   \tilde{\Phi}^K&=\frac{2C_{\mu} \mu}{(C_r r+1)^2+C_{\mu}^2\mu^2}.\label{SolK4}
\end{align}
However, they are not unique solutions of Ernst's equations (\ref{Eq1}) and (\ref{Ernst}). For any quadratic functions
\begin{align}
 \Delta^K_{r}&=r^2+2A r+B,  \\
  \Delta^K_{\mu}&=-(\mu^2+2C\mu+D),
\end{align}
where $A,B,C,D$ are real constants, we can obtain similar solutions. Defining
\begin{align}
  u & =\frac{r+A}{\sqrt{A^2-B}}, \\
  v & =\frac{\mu+C}{\sqrt{C^2-D}},
\end{align}
we have
\begin{align}
  \Delta^K_{r} &=(A^2-B)(u^2-1)\equiv(A^2-B)\Delta^K_u,  \\
  \Delta^K_{\mu} & =(C^2-D)(1-v^2)\equiv(C^2-D)\Delta^K_v.
\end{align}
Direct mathematical calculations show that $\Delta^K_u,\Delta^K_v,E$ also satisfy the Ernst's equations, i.e.
\begin{align}
  \Delta^K_{u,uu}+\Delta^K_{v,vv} & =0,\label{ger11} \\
  (1-EE^{*})[(\Delta^K_u E_{,u})_{,u}+(\Delta^K_{v}E_{,v})_{,v}]&=-2E^{*}[\Delta^K_uE_{,u}^2+\Delta^K_{v}E_{,v}^2].\label{ger22}
\end{align}
In virtue of these processes, we can obtain many new solutions of Ernst's equations.

Kerr metric is one of the above solutions, which is
\begin{align}
   \Delta^K_{r}&=r^2-2Mr+a^2,  \\
   \Delta^K_{\mu}&=1-\mu^2,\\
   E&=-C_r \frac{r-M}{\sqrt{M^2-a^2}}-i C_{\mu} \mu.
\end{align}
 The constants are chosen to be $C_r=\frac{\sqrt{M^2-a^2}}{M}$ and $C_{\mu}=\frac{a}{M}$. From $\tilde{Z}=-\frac{1+E}{1-E}=\tilde{\Psi}+i \tilde{\Phi}$, we have
\begin{align}
  \tilde{\Psi}^K & =\frac{\Delta^K_r-a^2\Delta^K_{\mu}}{r^2+a^2\mu^2},\label{SolKK1} \\
  \tilde{\Phi}^K & =\frac{2aM\mu}{r^2+a^2\mu^2}.
\end{align}
In the original metric, the unknown functions are
\begin{align}
\rho^2&=r^2+a^2\mu^2,\label{rhoK}\\
  \chi & =\frac{(r^2+a^2\mu^2)\sqrt{\Delta_r^{K}\Delta_{\mu}^{K}}}{(r^2+a^2)^2\Delta^K_{\mu}-a^2(1-\mu^2)^2\Delta^K_r},\label{chiK} \\
  q &=a\frac{(r^2+a^2)\Delta^K_{\mu}-(1-\mu^2)\Delta^K_r}{(r^2+a^2)^2\Delta^K_{\mu}-a^2(1-\mu^2)^2\Delta^K_r}.\label{qK}
\end{align}
If we define $\mu=\cos\theta$, the solutions transform to the standard Kerr metric form in the Boyer-Lindquist coordinate system, which is
\begin{equation}
  ds^2=-\frac{\rho^2\sin^2\theta\Delta^K_r}{\Sigma_K^2}dt^2+\frac{\Sigma_K^{2}}{\rho^2}\left(d\varphi-\frac{(r^2+a^2)- \Delta^K_{r}}{\Sigma_K^{2}}a\sin^2\theta dt\right)^2+\frac{\rho^2}{\Delta^K_r}dr^2+\rho^2d\theta^2,
\end{equation}
where
\begin{align}
  \rho^2 & =r^2+a^2\cos^2\theta, \\
  \Sigma_K^{2} & =(r^2+a^2)^2\sin^2\theta-a^2\sin^4\theta\Delta^K_{r},\\
  \Delta^K_{r}&=r^2-2Mr+a^2.
\end{align}

\section{The derivation of Kerr-Ads metric}
In the case of $R\neq 0$, we still expect the system of Eqs. (\ref{Eq1}), (\ref{Eq2}) and (\ref{Eq3}) to be closed. For this purpose, we try to presuppose the function $\rho^2$. The choice of $\rho^2$ is arbitrary. However, since $r$ and $\mu$ always appear in similar configurations, we can expect that $\rho^2$ should also have a similar configuration for $r$ and $\mu$. Notice that the left-hand side of Eq. (\ref{Eq1}) does not have terms $\Delta_{r}\Delta_{\mu}$ and $\Delta_{r,r}\Delta_{\mu,\mu}$, we can expect that term $r\mu$ is not included in $\rho^2$. One may try such terms as $r+\mu$, $r^2+\mu^2$, $r^3+\mu^3$.... Considering the form $\rho^2=r^2+a^2\mu^2$ in the case of $R=0$, we can set the function $\rho^2=r^2+\mu^2$ as the simplest choice.

The solution in the case of $R\neq 0$ may be obtained by the generalization of the solution in the case of $R=0$. In the spherical case $a=0$, it is not difficult to find that the Ricci scalar is a linear term in the metric. In the axisymmetric case $a\neq 0$, we expect this algebraic property is still true. Thus, we decompose the solutions into two parts
\begin{align}
  \Delta_{r} & = \Delta_{r}^{K}+R \Delta_{r}^{R},\label{Sol1}\\
  \Delta_{\mu} & = \Delta_{\mu}^{K}+R \Delta_{\mu}^{R},\label{Sol2}\\
  \tilde{\Phi}&=\tilde{\Phi}^{K}+R \tilde{\Phi}^{R},\label{Sol3}\\
  \tilde{\Psi}&=\tilde{\Psi}^{K}+R \tilde{\Psi}^{R},\label{Sol4}
\end{align}
where $\Delta_{r}^{R},\Delta_{\mu}^{R},\tilde{\Phi}^{R},\tilde{\Psi}^{R}$ are the solutions introduced by $R$. Notice that
\begin{equation}
  \Delta_{r}^{R}=\frac{\partial \Delta_{r}}{\partial R},~\Delta_{\mu}^{R}=\frac{\partial \Delta_{\mu}}{\partial R},~\tilde{\Phi}^{R}=\frac{\partial \tilde{\Phi}}{\partial R}, ~\tilde{\Psi}^{R}=\frac{\partial \tilde{\Psi}}{\partial R}.\nonumber
\end{equation}
Plugging the solutions (\ref{Sol1}) - (\ref{Sol4}) into Eqs. (\ref{Eq1}), (\ref{Eq2}) and (\ref{Eq3}), and taking the partial differential of $R$, we obtain
\begin{align}
 \Delta^R_{r,rr}+ \Delta^R_{\mu,\mu\mu} &=-(r^2+\mu^2),\label{StR1}\\
  \tilde{\Psi}^R[(\Delta^R_r \tilde{\Psi}^R_{,r})_{,r}+(\Delta^R_{\mu} \tilde{\Psi}^R_{,\mu})_{,\mu}]&= \Delta^R_r[(\tilde{\Psi}_{,r}^{R})^2-(\tilde{\Phi}_{,r}^{R})^2]+ \Delta^R_{\mu}[(\tilde{\Psi}_{,\mu}^{R})^2-(\tilde{\Phi}_{,\mu}^{R})^2]\nonumber\\
 &+\frac{(\tilde{\Psi}^{R})^2}{2}(\Delta^R_{r,rr}+\Delta^R_{\mu,\mu\mu}),\label{StR2}\\
 \tilde{\Psi}^R[(\Delta^R_r \tilde{\Phi}^R_{,r})_{,r}+(\Delta^R_{\mu} \tilde{\Phi}^R_{,\mu})_{,\mu}]&=2\Delta^R_r\tilde{\Phi}^R_{,r}\tilde{\Psi}^R_{,r}+2\Delta^R_{\mu}\tilde{\Phi}^R_{,\mu}\tilde{\Psi}^R_{,\mu}.\label{StR3}
\end{align}

To solve Eq. (\ref{StR1}), we let
\begin{equation}
 \Delta^R_{r,rr}=-(r^2+k_1^2),\quad \Delta^R_{\mu,\mu\mu}=-(\mu^2-k_1^2),
\end{equation}
where $k_1$ is a constant. Therefore, one can easily obtain the solution
\begin{equation}
  \Delta^R_{r}=-(\frac{r^4}{12}+\frac{k_1^2}{2}r^2),\quad \Delta^R_{\mu}=-\frac{\mu^4}{12}+\frac{k_1^2}{2}\mu^2.
\end{equation}
We still express $\tilde{\Psi}^R$ as the real part and $\tilde{\Phi}^R$ as the imaginary part of a complex function $Z$, i.e.
\begin{equation}\label{Zexp}
 Z=\tilde{\Psi}^R+i\tilde{\Phi}^R.
\end{equation}
Thus, Eqs. (\ref{StR2}) and (\ref{StR3}) can be expressed as
\begin{equation}\label{ErnstR3}
\mathrm{ Re}(Z)[(\Delta^R_r Z_{,r})_{,r}+(\Delta^R_{\mu} Z_{,\mu})_{,\mu}]=\Delta^R_rZ_{,r}^2+\Delta^R_{\mu}Z_{,\mu}^2-\frac{\mathrm{ Re}^2(Z)}{2}(r^2+\mu^2),
\end{equation}
where $\mathrm{ Re}^2(Z)=\mathrm{ Re}(Z)\cdot \mathrm{ Re}(Z)$. It's not hard to prove that Eq. (\ref{ErnstR3}) permits the solution
\begin{equation}
 Z=-\frac{1}{12}(r+i \mu)^2-\frac{k_1^2}{2}.
\end{equation}
From (\ref{Zexp}), we have
\begin{align}
   \tilde{\Psi}^R & =-\frac{1}{12}(r^2-\mu^2)-\frac{k_1^2}{2},\label{SoRR11} \\
  \tilde{\Phi}^R & =-\frac{r\mu}{6}.
\end{align}

The above solution also can be extended to the choice of $\rho^2=r^2+ k_2^2\mu^2$. Notice that the solution (\ref{SoRR11}) can also be written as
\begin{equation}
  \tilde{\Psi}^R=\frac{\Delta^R_{r}-\Delta^R_{\mu}}{r^2+\mu^2},
\end{equation}
coinciding with the solution (\ref{SolKK1}) where $a=1$. We then can introduce a new constant $k_2$ to the solution. Defining $\bar{\mu}=k_2\mu$ and replacing $\mu$ with $\bar{\mu}$ in Eqs. (\ref{Eq1}), (\ref{Eq2}) and (\ref{Eq3}), we know that $\Delta^R_{r}, \Delta^R_{\bar{\mu}}, \tilde{\Psi}^R(r,\bar{\mu}), \tilde{\Phi}^R(r,\bar{\mu})$ are also solutions. The function $\rho^2$ turns into $\rho^2=r^2+\bar{\mu}^2=r^2+ k_2^2\mu^2$. Since $\frac{\partial}{\partial \bar{\mu}}=\frac{\partial}{\partial \mu}\frac{\partial \mu}{\partial \bar{\mu}}=\frac{1}{k_2}\frac{\partial}{\partial \mu}$, Eqs. (\ref{Eq1}), (\ref{Eq2}) and (\ref{Eq3}) can be reproduced when letting $\Delta^R_{\mu}= \frac{1}{k_2^2}\Delta^R_{\bar{\mu}}$. Therefore, for $\rho^2=r^2+ k_2^2\mu^2$, Eqs. (\ref{Eq1}), (\ref{Eq2}) and (\ref{Eq3}) permit the solution
\begin{align}
  \Delta^R_{r}&=-(\frac{r^4}{12}+\frac{k_1^2}{2}r^2), \\
  \Delta^R_{\mu}&=-\frac{k_2^2\mu^4}{12}+\frac{k_1^2}{2}\mu^2,\\
     \tilde{\Psi}^R & =-\frac{1}{12}(r^2-k_2^2\mu^2)-\frac{k_1^2}{2},\label{SoRR1} \\
  \tilde{\Phi}^R & =-\frac{k_2r\mu}{6}.\label{SoRR2}
\end{align}

Till now, we already have the solutions both for $\Delta_{r}^{K},\Delta_{\mu}^{K},\tilde{\Phi}^{K},\tilde{\Psi}^{K}$ and $\Delta_{r}^{R},\Delta_{\mu}^{R},\tilde{\Phi}^{R},\tilde{\Psi}^{R}$. Combining them together and letting $k_1=k_2=a$, we obtain the final solution
\begin{align}
\Delta_{r}&=r^2-2M r+a^2-\frac{R}{12}r^2(r^2+a^2),\label{Soll1} \\
  \Delta_{\mu}&=1-\mu^2-\frac{R}{12}a^2\mu^2(\mu^2-1),\label{Soll2}\\
  \tilde{\Psi} & =\frac{\Delta_r-a^2\Delta_{\mu}}{r^2+a^2\mu^2},\label{Soll3}\\
  \tilde{\Phi} & =\frac{2M a \mu}{r^2+a^2\mu^2} -R\frac{ar\mu}{6}.\label{Soll4}
\end{align}
In the original metric, the unknown variables $\rho^2,\chi, q$ can also be expressed as
\begin{align}
\rho^2&=r^2+a^2\mu^2,\label{rhoSR}\\
  \chi & =\frac{(r^2+a^2\mu^2)\Delta_r^{\frac{1}{2}}\Delta_{\mu}^{\frac{1}{2}}}{(r^2+a^2)^2\Delta_{\mu}-a^2(1-\mu^2)^2\Delta_r},\label{chiSR} \\
  q &=a\frac{(r^2+a^2)\Delta_{\mu}-(1-\mu^2)\Delta_r}{(r^2+a^2)^2\Delta_{\mu}-a^2(1-\mu^2)^2\Delta_r}.\label{qSR}
\end{align}
Choosing $\mu=\cos\theta$, we can transform the solution into the Boyer-Lindquist coordinate
\begin{equation}
  ds^2=-\frac{\rho^2\Delta_r\Delta_{\theta}}{\Sigma^2}dt^2+\frac{\Sigma^2}{\rho^2}\left(d\varphi-\frac{(r^2+a^2)\Delta_{\theta}-\sin^2\theta \Delta_{r}}{\Sigma^2}adt\right)^2+\frac{\rho^2}{\Delta_r}dr^2+\frac{\rho^2\sin^2\theta}{\Delta_{\theta}}d\theta^2,
\end{equation}
where
\begin{align}
  \rho^2 & =r^2+a^2\cos^2\theta, \\
  \Sigma^2 & =(r^2+a^2)^2\Delta_{\theta}-a^2\sin^4\theta\Delta_{r},\\
  \Delta_{r}&=-\frac{R}{12}(r^4+a^2r^2)+r^2-2Mr+a^2,\\
  \Delta_{\theta}&=\sin^2\theta\left(1+\frac{R}{12}a^2\cos^2\theta\right).
\end{align}
This is the Kerr-Ads metric in a general general $f(R)$ theory.

\section{Conclusion and discussion}
As far as we know, at least three roads can reach rotating black holes in gravitational theories. The first is the Newman-Janis algorithm, which directly works on the solutions. The second is Carter's consideration of the spacetime's structure. The third is from Chandrasekhar, which is based on a strict analytical calculation. In this paper, we extend the Chandrasekhar's method into $f(R)$ theory. As an analytical calculation, the solving process is universal and can be established as a standard analytical calculation procedure to obtain the Kerr and Kerr-Ads metric, which is easy to be applied in other modified gravities.

We worked in a general $f(R)$ gravity theory and began with a general stationary axisymmetric metric, which have 5 unknowns and all of them depend on two variables. We use the moving frame to calculate the Einstein tensor. Suppose the spacetime is a 4-dimensional Riemannian manifold, which are fully described by RLC connection and curvature. Through calculating the Cartan's equation of structure, we extract the RLC connection forms and derive the curvature forms. Furthermore, we construct the Einstein tensor by using the curvature forms.

We chose the gauge freedom to transform the axisymmetric metric into a more symmetrical form, which can effectively reduce the difficulty of analytical calculation. Based on the symmetrical metric, we derived the field equations in $f(R)$ gravity. In order to give a simple solving process, we further transformed the variables in the field equations into the conjugate coordinates. At last, we solved them separately according to whether the Ricci scalar is zero.

In the case of $R=0$, the field equations reduce to the Ernst's equations. The Kerr metric can be obtained by generalizing the elementary solution of Ernst's equations. In the case of $R\neq 0$, the solution is decomposed into two parts: one is the Kerr solution and the second is introduced by $R$. Through some mathematical process, we analytically calculated the solution for the second part. Combining the two parts, we obtained the Kerr-Ads solution.

\textbf{Acknowledgement}
This work is partially supported by the National Natural Science Foundation of China (NSFC U2031112) and the Scientific Research Foundation of Hunan University of Arts and Sciences (E07023026). We also acknowledge the science research grants from the China Manned Space Project with NO. CMS-CSST-2021-A06.


\end{document}